\RequirePackage{fix-cm}
\documentclass[%
twocolumn%
,final%
]{svjour3}
\usepackage{graphicx}
\usepackage{bm}      

\usepackage{amssymb}

\begin{document}

\title{Dimesoatom breakup in the Coulomb field}

\author{L.G. Afanasyev\thanks{\email{Leonid.Afanasyev@cern.ch}} \and
S. R. Gevorkyan\thanks{\email{gevs@jinr.ru}}
\and O.O. Voskresenskaya\thanks{\email{voskr@jinr.ru}} }

\institute{Joint Institute for Nuclear Research, Dubna, Moscow Region, 141980 Russia}
\date{Received: \today / Revised version: September, 2019}

\maketitle

\begin{abstract}
Momentum and angular distributions of charged meson pairs $h^+h^-$ ($h=\pi, K$)  from elementary atoms (EA) breakup (ionization) in the Coulomb field of a target atom is considered in the Born and Glauber approximations.
Exploiting the fact that the atomic screening of the target Coulomb potential is important at small transfer momenta, while multi-photon exchanges are essential at large transfer momenta we express the cross sections
of EA breakup as a sum of two terms. In the region of modest transfer momenta the cross section is determined by the single-photon exchange (first Born approximation) accounting for the target atoms screening,
whereas at large transfer momenta using the unscreened potential allows to take into account all multi-photon exchanges and obtain the cross section of EA breakup in the close analytical form.


\end{abstract}

\section{Introduction}
The DIRAC experiment at Proton Synchrotron, CERN, have observed and studied  relativistic hydrogen-like elementary atoms (EA) formed by  pairs of $\pi^+\pi^-$ and  $\pi^\pm K^\mp$ mesons
($A_{2\pi}$  and $A_{\pi K}$ atoms, respectively).
Determination of the ground-state lifetime of these atoms allowed one to obtain the difference of $\pi\pi$ ($a_0-a_2$) and $\pi K$ ($a_{1/2}-a_{3/2}$) scattering lengths in the S state with isospin $I=0,2$ and $I=1/2,\,3/2$ correspondingly in a model independent way \cite{AFAN93,ADEV11,ADEV17}.
The method of EA observation proposed in \cite{NEME85} is based on detection of oppositely charged meson pairs $h^+h^-$ ($h=\pi, K$) from EA breakup (ionization) in the Coulomb field of a target atom. Total cross sections of EA with target atoms as well as cross sections of transition between discrete states of EA and their breakup (ionization) were calculated in set of works \cite{AFAN96,AFAN04,AFAN03,BASEl00,BASEl01,BASEl02} and were used in calculations the EA lifetimes in the DIRAC experiment.
A distribution of pairs from atoms breakup (atomic pairs) on their relative momentum in the rest frame has a scale of atomic Born momentum which is extremely small compared to haronic processes. This property was used to detect such pairs over the huge background of pairs produced directly in hadronic processes. In the current paper the relative momentum and angular distributions of atomic pairs are obtained in different approaches in order to confirm the approach used in the DIRAC experiment. The first attempts for numerical estimation of the breakup spectra of EA in different approaches were reported in \cite{HadAt01}. Here we obtained the general analytical expressions and consider the different approximations with the relevant calculations for the breakup spectra.

\section{Dipole approach}

The simplest way to get the relative momentum spectra of atomic pairs from EA breakup in the Coulomb field of target atoms is to use, so called, dipole approach in the first Born approximation. It was published in PhD Thesis  \cite{afan97} only, so here we present a short derivation of formulas to compare with more accurate approaches.

The cross section of $\pi^+\pi^-$ atoms with target atoms in the first Born approximation \cite{AFAN96} is written as:
\begin{eqnarray}\label{eq:d1}
&&\!\! \frac{d\sigma_i}{d^3p}(p,\theta) = 8\pi {\alpha^2 \over \beta^2}
\! \int\limits_0^\infty \! {dq\over q^3}\,| F_A(q)|^2 \times \\
&& \qquad\times \left|S_{\vec{p},i}(\vec{p},\vec{q}/2) - S_{\vec{p},i}(\vec{p},-\vec{q}/2)\right|^2 \,.
\nonumber
\end{eqnarray}
Here $\vec{p}$ is the relative momentum of EA components in their rest frame, $\theta$ is the polar angle relative to the EA full momentum, $\alpha$ is the fine structure constant, $\beta=v/c$ is the EA velocity, $q$ is transfer momentum, $F_A(q)$ is the elastic form factor of the target atom, $S_{\vec{p},i}(\vec{p},\vec{q})$ is the EA transition form factor from the initial state $i$ to the continues final state $\vec{p}$ which is expressed via EA wave functions as:
\begin{equation}\label{eq:dff}
S_{\vec{p},i}(\vec{p},\vec{q})=\int \! d\vec{r}\,\exp{(i\vec{q}\vec{r})}
\psi_f^*(\vec{r})\,\psi_i(\vec{r}) \,.
\end{equation}
Substituting (\ref{eq:dff}) in (\ref{eq:d1}) we get the cross section as:
\begin{eqnarray}\label{eq:d2}
&&\!\! \frac{d\sigma_i}{d^3p}(p,\theta) = 8\pi {\alpha^2 \over \beta^2}
\! \int\limits_0^\infty \! {dq\over q^3}\,| F_A(q)|^2 \times \\
&& \qquad\times \left|{\int\! d\vec{r} \,(\exp{( i\vec{q}\vec{r}/2)} -
                          \exp{(-i\vec{q}\vec{r}/2)})
\psi_{\vec{p}}^* (\vec{r})\,\psi_i(\vec{r}) }\right|^2.
\nonumber
\end{eqnarray}
The product $\vec{q}\vec{r}$ in the exponents are of the order $\lesssim5\cdot10^{-2}$ because of $\vec{q}$ is limited by $F_A(q)$ and $\vec{r}$ limited by the size of initial state $\psi_i(\vec{r})$. Thus the exponents can be replaced by the linear expansion, so called dipole approach, and the cross section is expressed via two independent integrals:
\begin{equation}\label{eq:d3}
\frac{d\sigma_i}{d^3p}(p,\theta) = 8\pi {\alpha^2 \over \beta^2}
\! \int\limits_0^\infty \! {dq\over q^3}\,| F_A(q)|^2
\left|{\int\! d\vec{r} \, (\vec{e}\vec{r})
\psi_{\vec{p}}^* (\vec{r})\,\psi_i(\vec{r}) }\right|^2,
\end{equation}
where $\vec{e}=\vec{q}/|\vec{q}|$ is the unit vector. In this approximation the first integral
divergent logarithmically and requires truncation at a high $q$, whereas the second integral provides the required distribution. Thus, in the dipole approach the relative momentum spectrum of atomic pairs is fully dictated by the EA wave functions. Using the pure Coulomb wave function (see discussion in \cite{amirh99}) for the ground state ($1S$) we have:
\begin{equation}\label{eq:d1s}
\frac{d\sigma_{1S}}{dp\,\sin\theta d\theta} = C \:
\frac{x\, \exp{\left({-4 x^{-1}\arctan{x}}\right)}}%
{(x^2+1)^5 \,\left(1-\exp{(2\pi/x)}\right)} \:\sin^2{\theta},
\end{equation}
where $x$ is express via EA Bohr momentum $p_B$ as $x=p/2p_B$ and $C$ is the normalization constant. For the initial state $2S$ the correspondent distribution is written as:

\begin{equation}\label{eq:d2s}
\frac{d\sigma_{2S}}{dp\,\sin\theta d\theta} = C' \:
\frac{x (x^2+1) \exp{\left({-4x^{-1}\arctan{2x}}\right)}
}%
{(4x^2+1)^6 \,\left(1-\exp{(2\pi/x)}\right)}\: \sin^2{\theta}.
\end{equation}

\section{Spectra of mesons from the EA breakup}

To obtain the spectra of mesons from EA ionization in the Coulomb field of the target atoms, we exploit the fact that the atomic screening is essential at small transfer momenta (large impact parameters),
while the multiple photon exchanges are significant at large transfer momenta (small impact parameters) where one can safely neglect the atomic screening.
Keeping in mind this observation let us represent the differential cross section for EA  breakup in the Coulomb field of the target as a sum of two terms: the first one corresponds to the Born amplitude (single photon exchange) accounting for atomic screening in the Coulomb field of the target, whereas the second term accounts to all multiple exchanges of photons in unscreened Coulomb potential:
\begin{eqnarray}
\frac{d\sigma}{d^3p}(p,\theta) & = & \frac{1}{(2\pi)^5}\int_{0}^{q_0}qdqd\varphi\vert A_{fi}^B(\vec p,\vec q)\vert^{2}+ \nonumber\\
&&+\frac{1}{(2\pi)^5}\int_{q_0}^{\infty}qdqd\varphi\vert A_{fi}^G(\vec p,\vec q)\vert^2.
\end{eqnarray}
Here $\vec p$ is the mesons relative momentum  in the final state in their rest frame and $\vec q=\vec k-\vec k'$ the two-dimensional transfer momentum. Remembering that the inverse screening radius of target atoms
 $\lambda^{-1}\approx m_e\alpha Z^{1/3}$ (Z atomic number;  $m_e$ electron mass), while the EA Bohr momentum is $\mu\alpha$ ($\mu={m_1m_2}/(m_1+m_2)$ EA reduced mass),
 the boundary momentum we choose as  $q_0\sim \alpha (m_e\mu Z^{1/3})^{1/2}$.

The general form of the relativistic EA ionization amplitude can be written in the eikonal   approximation~\cite{gtv98,agv17}
\begin{eqnarray}
A_{fi}(\vec p,\vec q)&=&\frac{i}{2\pi}\int e^{i\vec q\vec b}d^2b d^3r\psi_f^{\ast}(\vec p,\vec r)\psi_i(\vec r) \left[1-e^{i\Delta\chi(\vec b,\vec s)}\right]\nonumber\\
\label{eq:2}
&=&\int d^2s f(\vec q,\vec s)h_{fi}(\vec p,\vec s),\\
f(\vec q,\vec s)&=&\frac{i}{2\pi}\int d^2b\left[1-e^{i\Delta\chi(\vec b,\vec s)}\right]e^{i\vec q\vec b},\\
\Delta\chi(\vec b,\vec s)&=& \chi (\vec b-\vec s/2)-\chi (\vec b+\vec s/2),
\nonumber\\
\chi (\vec b)&=&\frac{1}{\beta}\int U(\vec b,\xi)d\xi,\nonumber\\
\label{eq:4}
h_{fi}(\vec p,\vec s)&=&\int\limits_{-\infty}^{\infty}dz\psi_f^{\ast}(\vec p,\vec r)\psi_i(\vec r)\,, \quad \vec r=(\vec s,z).
\end{eqnarray}
Here $\vec s$  is the projection of the  distance between EA constituents  $\vec r$ on the plane of the impact parameter b.
The phase shift $\chi (\vec b)$ is expressed via the Coulomb potential of the target atom U(r) in the standard way.\\
The general form of  EA wave function  with  n,l,m the principal, angular and magnetic quantum numbers  reads~\cite{sommer,land}
\begin{eqnarray}
\psi_i(\vec r)&=&\psi_{nlm}(\vec r)= Y_{lm}\left(\frac{\vec r}{r}\right)R_{nl}(r)
\end{eqnarray}
with the radial part
\begin{eqnarray}
&&R_{nl}(r)=\frac{2\omega^{\frac{3}{2}}}{\Gamma(2l+2)}
\left[\frac{\Gamma(n+l+1)}{n\Gamma(n-l)}\right]^{\frac{1}{2}}(2\omega r)^l
\times\nonumber\\
&&\times\Phi(-n+l+1,2l+2;2\omega r)\times
\exp(-\omega r)\\
&&=2\omega^{\frac{3}{2}}
\left[\frac{\Gamma(n-l)}{n\Gamma(n+l+1)}\right]^{\frac{1}{2}}(2\omega r)^l
L^{2l+1}_{n-l-1}(2\omega r)\cdot
\exp(-\omega r)\nonumber.
\end{eqnarray}
 Here, $\omega=\mu\alpha/n$, $\Phi(a,b,c) $ is the confluent hypergeometric function and $L_{\beta}^{\alpha}(x)$ are  associated Laguerre polynomials.
The wave function of  the final (continuum) state has the form  \cite{land}
\begin{eqnarray}\label{eq:7}
\psi_f(\vec p,\vec r)&=& c^{(-)}\exp(i\vec p\vec r)
\cdot\Phi\left[-i\xi,1,-i(pr+\vec p\vec r)\right]\,,\\
c^{(-)}&=&(2\pi)^{-\frac{3}{2}}\exp\left(\frac{\pi\xi}{2}\right)
\Gamma(1+i\xi)\,, \quad \xi=\mu\alpha/p. \nonumber
\end{eqnarray}
Now we can proceed further considering Born and Glauber approximations separately.

\section{Born approximation}
In Born approximation the differential cross section of EA breakup in the Coulomb field of nuclei is expressed through the relevant form factors.
Expanding  the exponent in (\ref{eq:2}) and confining  by the first order in $\chi$  leads to the double differential cross section
of EA ionization in the following form:
\begin{eqnarray}
\label{eq:8}
\frac{d^3\sigma^{}_\textrm{B}}{d^3\vec  pd^2 q}&=&
\frac{1}{(2\pi)^3} U^2(\vec q)\left (S_{\vec p,nlm}(\vec  p,\vec  q_1 )- S_{\vec p,nlm}(\vec  p,\vec  q_2 )\right )^2,\nonumber\\
\vec q_1&=&\frac{m_1}{m_1+m_2}\vec q , \quad \vec q_2=-\frac{m_2}{m_1+m_2}\vec q.
\end{eqnarray}
The  transition form factor from any discrete to continuum state  defined by the equation
\begin{equation}
S_{fi}(\vec q,\vec p)=S_{\vec p,nlm}(\vec q,\vec p)=\int \psi_f^{\ast}(\vec r)e^{i\vec q\vec r} \psi_i(\vec r)d^3r.
\end{equation}
Usually~\cite{omidvar69,barut89} in ionization form factors  calculation the wave function of the final state  expands in  series of spherical harmonics reducing the problem to calculation
of infinite number of transition form factors. In this approach only the finite number of terms can be taken into account, leading to necessity  to estimate the error which one admit acting in such a  way.

On the other hand the ionization form factor for any discrete initial states can be represented in the closed analytical form as a finite sum of  terms expressed through
the classical  Gegenbauer and Jacobi  polynomials  $C^{\lambda}_n(x), P^{\alpha,\beta}_n(x)$ and thus can be evaluated numerically  with arbitrary  degree of accuracy (for details see Appendix).
 In the final form they reads:
\begin{eqnarray}
\label{eq:10}
&&\!\! S_{\vec p,nlm}(\vec q,\vec p)=\frac{4\pi\omega^{1/2}}{\omega^2+\Delta^2}\left(\frac{4i\omega p}{\omega^2+\Delta^2}\right)^l
\left[\frac{\Gamma(n-l)}{n\Gamma(n+l+1)}\right]^{\frac{1}{2}}
\times\nonumber\\
&&\quad\times\sum_{s=0}^{l}G_{lms}(\vec p,\vec q)H_{nls}(\vec p,\vec q)\left(\frac{\omega^2+\Delta^2}{(\omega-ip)^2+q^2}\right )^{i\xi+s},\\
&&G_{lms}(\vec q,\vec p)=(-1)^s\frac{\Gamma(i\xi+s)}{\Gamma(i\xi-l+s)\Gamma(s+1)}\times\nonumber\\
&&\qquad\times\sum_{l_1=s}^{l}\left(-\frac{q}{p}\right)^{l_1}\left[\frac{4\pi\Gamma(2l+2)}{\Gamma(2l_1+2)\Gamma(2l-2l_1+2)}\right]^{\frac{1}{2}}\times\nonumber\\
&&\qquad\quad\times\frac{\Gamma(l_1+1)}{\Gamma(l_1-s+1)}
\times\left[Y_{l_1}\left(\frac{\vec q}{q}\right)\otimes
Y_{l-l_1}\left(\frac{\vec p}{p}\right)\right]_{lm}\, ,\nonumber\\
&&H_{nls}(\vec p,\vec q)=(n+l)F_{n_1ls}(\vec p,\vec
q) - (n-l)F_{n_2ls}(\vec p,\vec q);\nonumber\\
&&n_1=n-l-1\, ,\quad n_2=n-l-2\, ,\nonumber\\
&&F_{n_{1(2)}ls}(\vec p,\vec q)=
\frac{\Gamma(l-s+\frac{1}{2}-i\xi)}{\Gamma(2l-2s+1-2i\xi)}\times\nonumber\\
&&\qquad\times\sum_{k=0}^{n_{1(2)}}w^kC_k^{(i\xi+s)}(v)
\frac{\Gamma(n_{1(2)}-k+2l-2s+1-2i\xi)}{\Gamma(n_{1(2)}-k+l-s+\frac{1}{2}-i\xi)}
\times\nonumber\\
&&\qquad\quad\times
P_{n_{1(2)}-k}^{(l-s-\frac{1}{2}-i\xi,l-s+\frac{1}{2}-i\xi)}(u) \, .\nonumber
\end{eqnarray}
Here $\bm{\Delta}=\vec q-\vec p\;$,
$u=(\Delta^2-\omega^2)/(\Delta^2+\omega^2)$,
$v=(q^2-p^2-\omega^2)/\sqrt{\left[(\omega-ip)^2+q^2\right]/\left[(\omega+ip)^2+q^2\right]}$,
$w=\sqrt{\left[(\omega+ip)^2+q^2\right]/\left[(\omega-ip)^2+q^2\right]}$.

Making use equations (\ref{eq:8}) and (\ref{eq:10}) we calculated the double differential cross section for EA transition from initial $S$ states with principal quantum numbers from $n=1$ upto $n=10$ to continuous spectra in the screened Coulomb field for the Ni target. For the numerical calculations we used the Moliere  parametrization of the Thomas-Fermi  potential energy
\begin{eqnarray}
&&U(q)= 4\pi Z\alpha \left (\frac{0.35}{q^2+\beta_1^2}+\frac{0.55}{q^2+\beta_2^2}+\frac{0.10}{q^2+\beta_3^2}\right ),\nonumber\\
&&\qquad\beta_1=\frac{0.3 Z^{1/3}}{0.885a_0},\quad \beta_2=4\beta_1,\quad\beta_3=5\beta_2, \nonumber\\
&&\qquad a_0=0.529\times 10^{-8}\mathrm{cm}.
\end{eqnarray}

\begin{figure}[h]
\begin{center}
\includegraphics[width=\columnwidth]{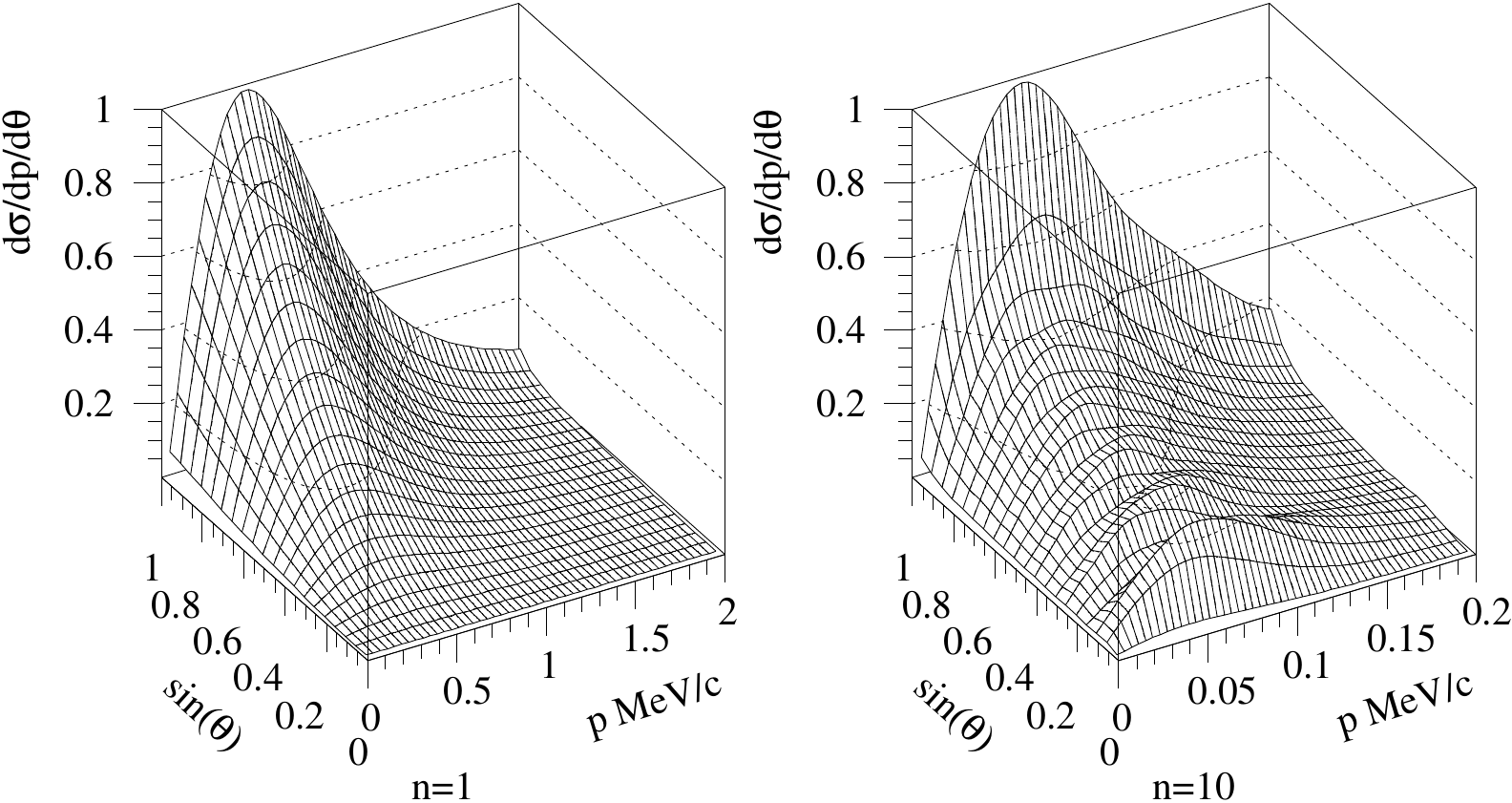}%
\caption{Momentum and angular distributions of $\pi^+\pi^-$ atomic pairs for ionization (breakup) from $1S$ and $10S$ states.}
\label{fig:1s-10s}
\end{center}
\end{figure}

\vspace{1mm}
The result of such calculations is illustrated in Fig.~\ref{fig:1s-10s} for $1S$ and $10S$ as 2D plots.

\vspace{3mm}

\section{Multiple exchanges}
The amplitude of two pions dipole with transverse size $\vec s$ scattered  with  momentum transfer $\vec q$  on the Coulomb field with  all multiple photon
exchanges taking into  account reads:
\begin{eqnarray}
\label{eq:12}
&&f(\vec q,\vec s)=\frac{i}{2\pi}\int d^2b\Gamma(\vec b,\vec s)e^{i\vec q\vec b};\;\;
\Gamma(\vec b,\vec s)=1-e^{i\Delta\chi(\vec b,\vec s)};\nonumber\\
&&\Delta\chi(\vec b,\vec s)= \chi (\vec b-\vec s/2)-\chi (\vec b+\vec s/2);
\nonumber\\
&& \chi (\vec b)=\frac{1}{\beta}\int U(\vec b,z)dz\;.
\end{eqnarray}
As was mentioned above the multiple exchanges are essential at small impact parameters, where one can safely used unscreened Coulomb potential.
In this case the Coulomb phase difference  takes the form~\cite{gtv98}:
\begin{eqnarray}
\Delta\chi(\vec b, \vec s)&=&-\nu\ln\left(\frac{b^2+\vec b\vec s+s^2/4}{b^2-\vec b\vec s+s^2/4}\right); \quad\nu=Z\alpha/\beta.
\end{eqnarray}
To proceed further let us employ the relation

\begin{eqnarray}
1-X^Y=-X Y F(1-Y, 1,2, 1-X)
\end{eqnarray}
and the integral representation of hypergeometric function
\begin{eqnarray}
&&F(1-Y,1,2,1-X)=
\nonumber\\
&&\qquad\frac{1}{\Gamma(1-Y)\Gamma(1+Y)}\int dt\frac{t^{-Y}(1-t)^Y}{1-t(1-X)}.
\end{eqnarray}
Making use this relations one gets
\begin{eqnarray}
\label{eq:16}
&&\Gamma(\vec b,\vec s))= 1-e^{i\Delta\chi(\vec b,\vec s)}=1-\left(\frac{b^2+\vec b\vec s+s^2/4}{b^2-\vec b\vec s+s^2/4}\right)^{i\nu}\nonumber\\
&&\qquad =- \frac{1}{\Gamma(1-i\nu)\Gamma(1+i\nu)}\int\limits_0^1 dx\left(\frac{x}{1-x}\right)^{i\nu} \times\nonumber\\
&&\qquad\quad\times\frac{2\vec b\vec s}{b^2-\vec b\vec s(2x-1)+s^2/4}\nonumber\\
&&\qquad =-\frac{1}{\Gamma(1-i\nu)\Gamma(1+i\nu)}
\int\limits_{-\infty}^{\infty} d\omega\frac{e^{2i\nu\omega}}{\cosh^2\,\omega} \times\nonumber\\
&&\qquad\quad\times\frac{\vec b\vec s}{b^2-\vec b\vec s\cdot
\tanh\omega+s^2/4}.
\end{eqnarray}
The  last expression is a result of the  replacement $ 2x-1\to\tanh \omega$.

 Substituting this expression in (\ref{eq:12}) one obtains the amplitude of dipole elastic scattering of the unscreened Coulomb field accounting to all multiple exchanges.

In calculating the overlap of wave function $h_{fi}(\vec p,\vec s)$
(\ref{eq:4}) we confined by the breakup of the  EA ground state for which the wave function reads
\begin{eqnarray}
\label{eq:17}
\psi_{i}(\vec r)=\frac{(\mu\alpha)^{3/2}}{\sqrt{\pi}}\exp(-\mu\alpha r).
\end{eqnarray}

Using the integral representation for the confluent hypergeometric functions
\begin{eqnarray}
\Phi\left(\alpha,1;w\right)=-\frac{1}{2\pi i}
\!\oint\limits_{C}^{}\!dt (-t)^{\alpha-1}(1-t)^{-\alpha}\exp(wt)dt
\end{eqnarray}
and the relation for the zero-order MacDonald function $K_0$
\begin{equation}
\int\limits_{-\infty}^{\infty}\!\!\frac{dz}{\sqrt{z^2+s^2}}
\exp[-a\sqrt{z^2+s^2}-bz]=2K_0\left(\sqrt{a^2-b^2}s\right)
\end{equation}
and also substituting the expressions for initial (\ref{eq:17}) and final (\ref{eq:7}) wave functions in expression (\ref{eq:4}), we get
\begin{eqnarray}
\label{eq:20}
&&\!\!\!\! h_{fi}(s)=\int \psi^{\ast}_{f}(\vec r)\psi_{i}(\vec r)dz\nonumber\\
&&=i\left(\frac{\mu\alpha}{\pi}\right)^{3/2}\!\!\!c^{(-)}
\frac{\partial}{\partial\mu\alpha}\frac{\partial}{\partial c}
\oint\limits_{C}^{}(-t)^{i\xi-1}(1-t)^{-i\xi}cK_1(cs)dt;\nonumber\\
&&c=\sqrt{(\mu\alpha-ipt)^2+p^2_L(1-t)^2}.
\end{eqnarray}
Using the expressions (\ref{eq:16}) and (\ref{eq:20}) one can carry out the  integration in (\ref{eq:4}) with the result
\begin{equation}
A_{fi}(\vec q)=i\left(\frac{\mu\alpha}{\pi}\right)^{3/2}\!\!\!\!c^{(-)}
\!\!\!\oint\limits_{C}^{}dt(-t)^{i\xi-1}\!(1-t)^{-i\xi}\!A_{fi}(\vec q,t),
\label{eq:21}
\end{equation}
\begin{eqnarray}
A_{fi}(\vec q,t)&=&\frac{\partial}{\partial\mu\alpha}
\frac{1}{c}\frac{\partial}{\partial c}
\int\limits_{-\infty}^{\infty}\frac{d\omega\; e^{2i\nu\omega}}
{\cosh\omega\vert\Gamma(i\nu)\vert^2}\times \nonumber\\
&& \times\int\limits_{0}^{\infty}dv\frac{\cosh v\cdot c}
{q[c^2+d^2+e^2+2cd \cosh v]}\,,
\end{eqnarray}
\begin{equation}
\label{eq:638}
d=\frac{q}{2 \cosh\omega},\
e^2=\left(\frac{\vec q}{2} \tanh\,\omega-\vec \kappa\right)^2,\
\vec \kappa=\vec p_T(1-t)\,.
\end{equation}
Changing the integration variable $\omega=\omega'+\omega_0$ by $\omega_0$ which is determined from the relations:
\begin{equation}
\cosh\omega(c^2+q^2/4+\kappa^2)=f \cosh\omega_0,\
\vec q\vec\kappa sh\,\omega=f \sinh\omega_0,
\end{equation}
we get
\begin{eqnarray}
&&\!\!\!\!\! A_{fi}(\vec q,t)=\frac{\partial}{\partial\mu\alpha}
\left(\frac{1}{c} \frac{\partial}{\partial c}\right)\\
&&\left\{\frac{e^{2i\nu\omega_0}}{\Gamma(i\nu)^2}
\int\limits_{-\infty}^{\infty}
\!\!d\omega'\; e^{2i\nu\omega'}
\!\int\limits_{0}^{\infty}\!\!dv\frac{ \cosh v\cdot c}
{[f \cosh\omega'+2cq\cdot \cosh v]}\right\}, \nonumber
\end{eqnarray}
\begin{eqnarray}
f^2&=&(c^2+q^2/4+\kappa)^2-(\vec q\vec\kappa)^2 \nonumber\\
&=&\left[c^2+(q/2+\kappa)^2\right]\cdot\left[c^2+(q/2-\kappa)^2\right]
\end{eqnarray}
with
\begin{eqnarray}
\label{eq:6312}
e^{2i\nu\omega_0}=\left[\frac{c^2+(q/2+\kappa)^2}
{c^2+(q/2-\kappa)^2}\right]^{i\nu}\,.
\end{eqnarray}
As a result we get
\begin{eqnarray}\label{eq:28}
&&A_{fi}(\vec q,t) = \frac{\vert\Gamma(1+i\nu)\vert^2}{q^2}\frac{\partial}{\partial\mu\alpha}
\left(\frac{1}{c}\frac{\partial}{\partial c}\right)
\nonumber\\
&&\quad\left\{\left[\frac{c^2+(q/2+\kappa)^2}
{c^2+(q/2-\kappa)^2}\right]^{i\nu}\!\!F\!\!\left(i\nu,-i\nu;1;1-
\frac{c^2q^2}{f^2}\right)\!\!
\right\}.
\end{eqnarray}
Expressions (\ref{eq:21}), (\ref{eq:28}) are the main result of our work. First time we obtain the close analytical expression for the ionization amplitude of ground
state elementary atom in the Coulomb field accounting for all multiple   exchanges of EA  in its interaction with the target as well as all Coulomb interactions between two mesons in initial and final states.

\section{Comparison and results}

\begin{figure}
\begin{center}
\includegraphics[width=\columnwidth]{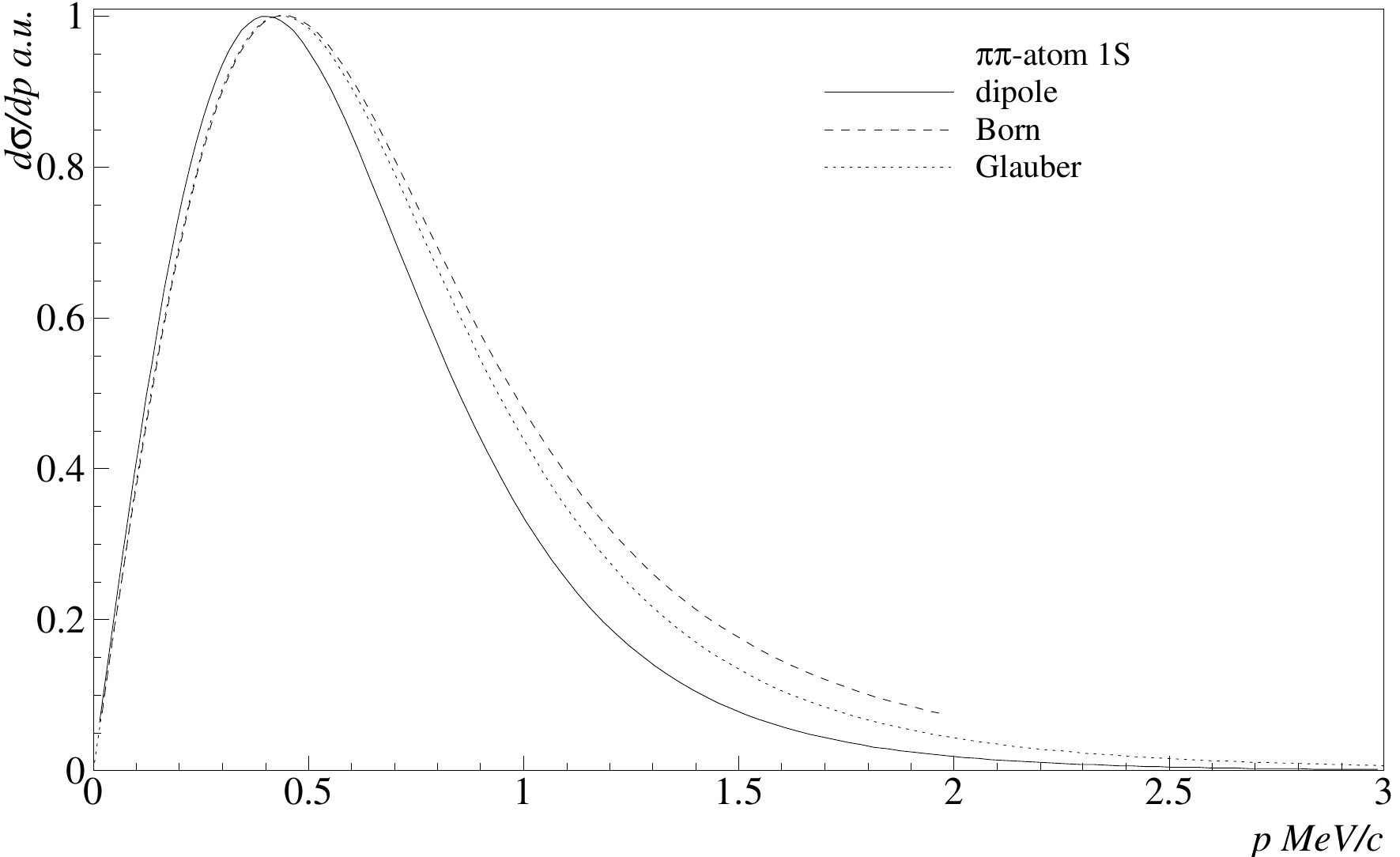}%
\caption{Relative momentum distributions of pions from $1S$ state $\pi^+\pi^-$ atom breakup in different approaches: solid line for the dipole approach (\ref{eq:d1s}), dashed line for the Born approximation and dotted line for the Glauber approximation.
\label{fig:q1s}}
\end{center}
\end{figure}

\begin{figure}
\begin{center}
\includegraphics[width=\columnwidth]{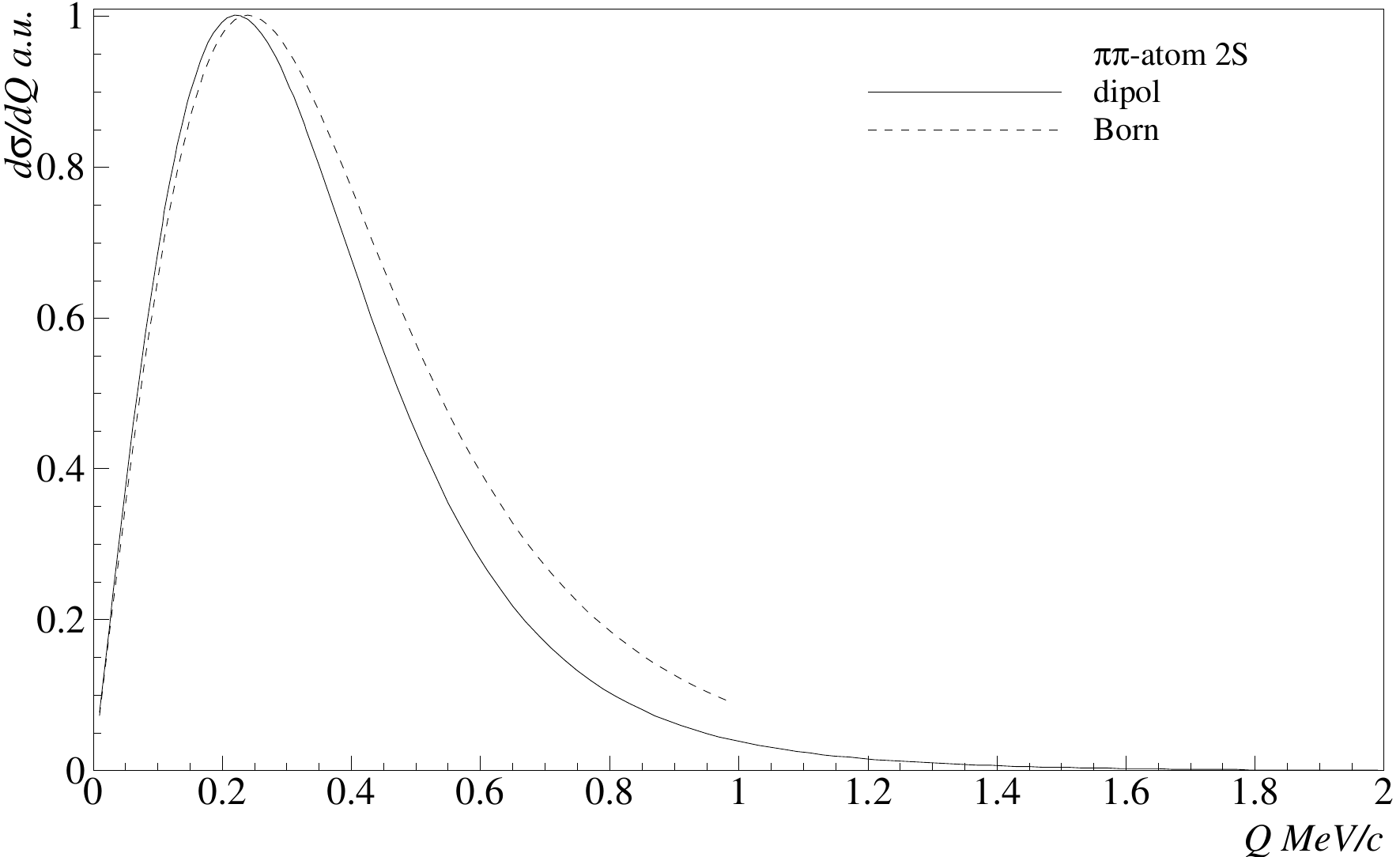}%
\caption{Relative momentum distributions of pions from $2S$ state $\pi^+\pi^-$ atom breakup in different approaches: solid line for the dipole approach (\ref{eq:d2s}), dashed line for the Born approximation and dotted line for Glauber approximation.
section for $A_{2\pi}^{2S}$.
\label{fig:q2s}}
\end{center}
\end{figure}

\begin{figure}
\begin{center}
\includegraphics[width=\columnwidth]{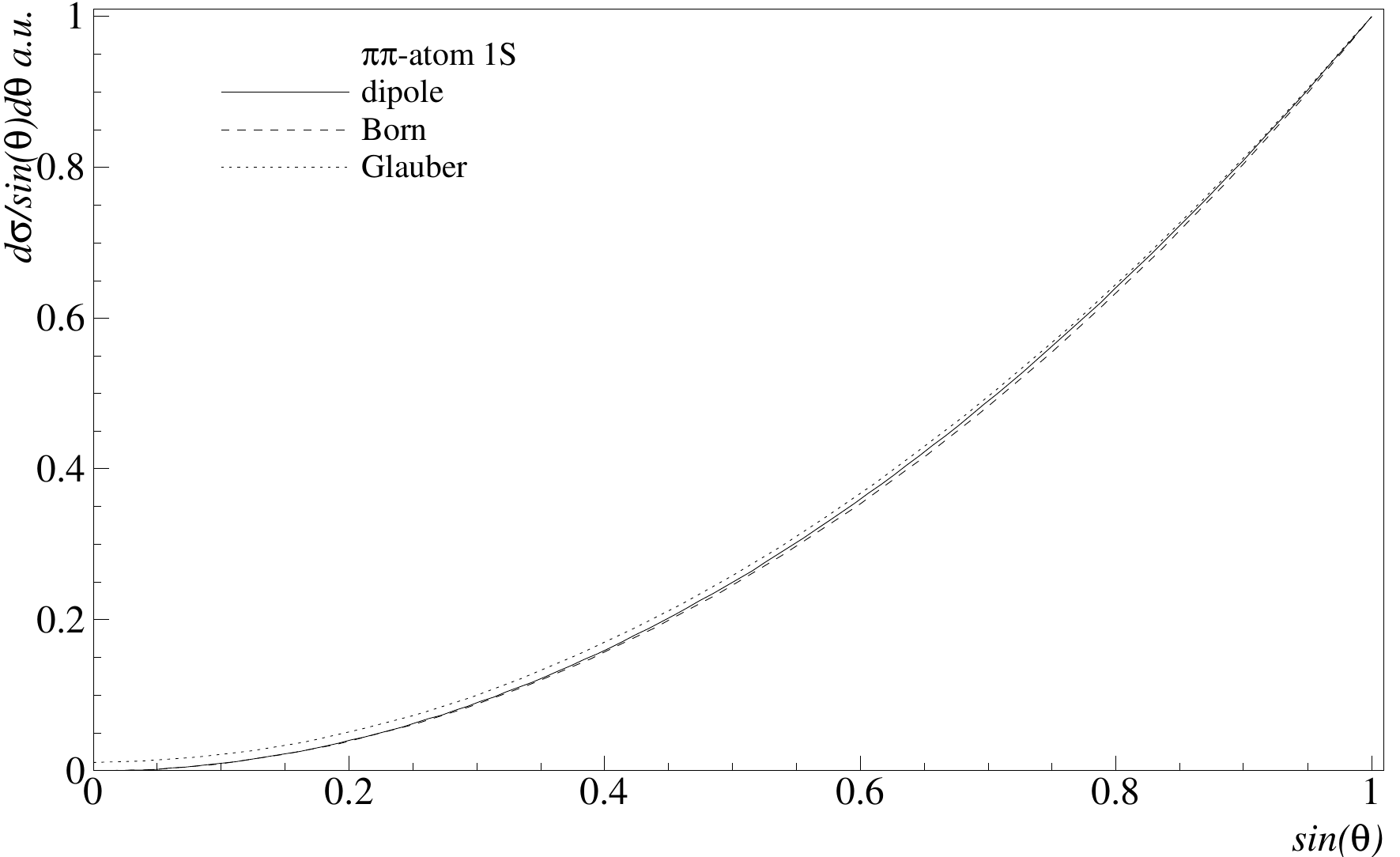}%
\caption{Angular distributions of pions from $1S$ state $\pi^+\pi^-$ atom breakup in different approaches: solid line for the dipole approach (\ref{eq:d1s}), dashed line for the Born approximation and dotted line for the  Glauber approximation.
\label{fig:t1s}}
\end{center}
\end{figure}

\begin{figure}
\begin{center}
\includegraphics[width=\columnwidth]{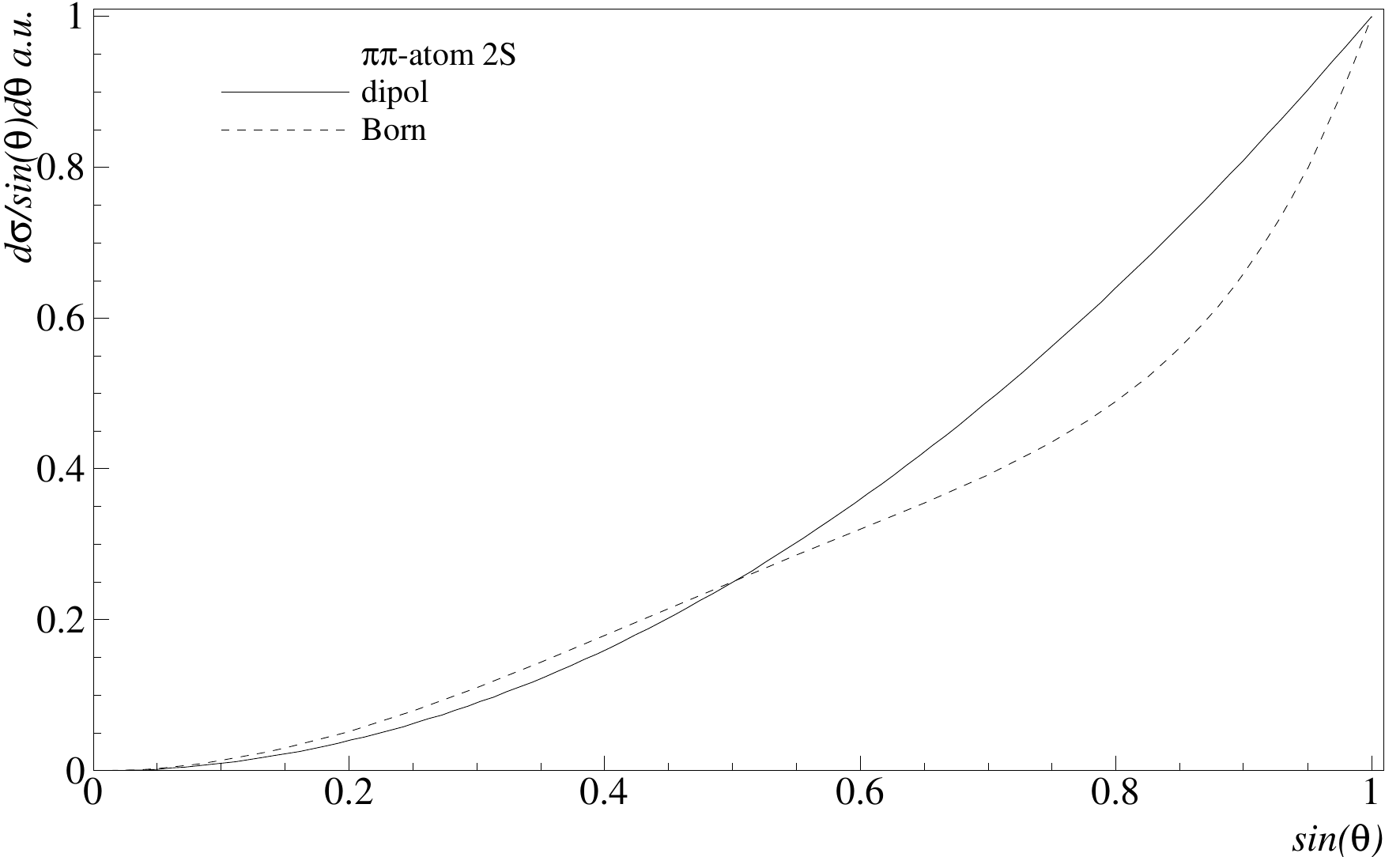}%
\caption{Angular distributions of pions from $2S$ state $\pi^+\pi^-$ atom breakup in different approaches: solid line for the dipole approach (\ref{eq:d2s}), dashed line for the Born approximation and dotted line for Glauber approximation.
\label{fig:t2s}}
\end{center}
\end{figure}

Let us compare the spectra of atomic pairs calculated in the different approaches mentioned above. Figures \ref{fig:q1s} and \ref{fig:q2s} show the relative momentum distributions of $\pi^+\pi^-$ atomic pairs calculated for $1S$ and $2S$ initial states. Figures \ref{fig:t1s} and \ref{fig:t2s} show the angular distributions of $\pi^+\pi^-$ atomic pairs calculated for $1S$ and $2S$ initial states. It is significant that even the simplest dipole approach gives the spectra very similar to ones calculated in more sophisticated approaches. Obviously, that this conclusion can be done only after all above calculations.

It is worth mentioning that the peak position in the relative momentum spectra are closed to the mean momentum of the initial atomic state. Thus the spectra become more narrow with increase of principal quantum number $n$ (see also Fig.~\ref{fig:1s-10s}), that is important for analysis of experimental data.

\begin{figure}
\begin{center}
\includegraphics[width=\columnwidth]{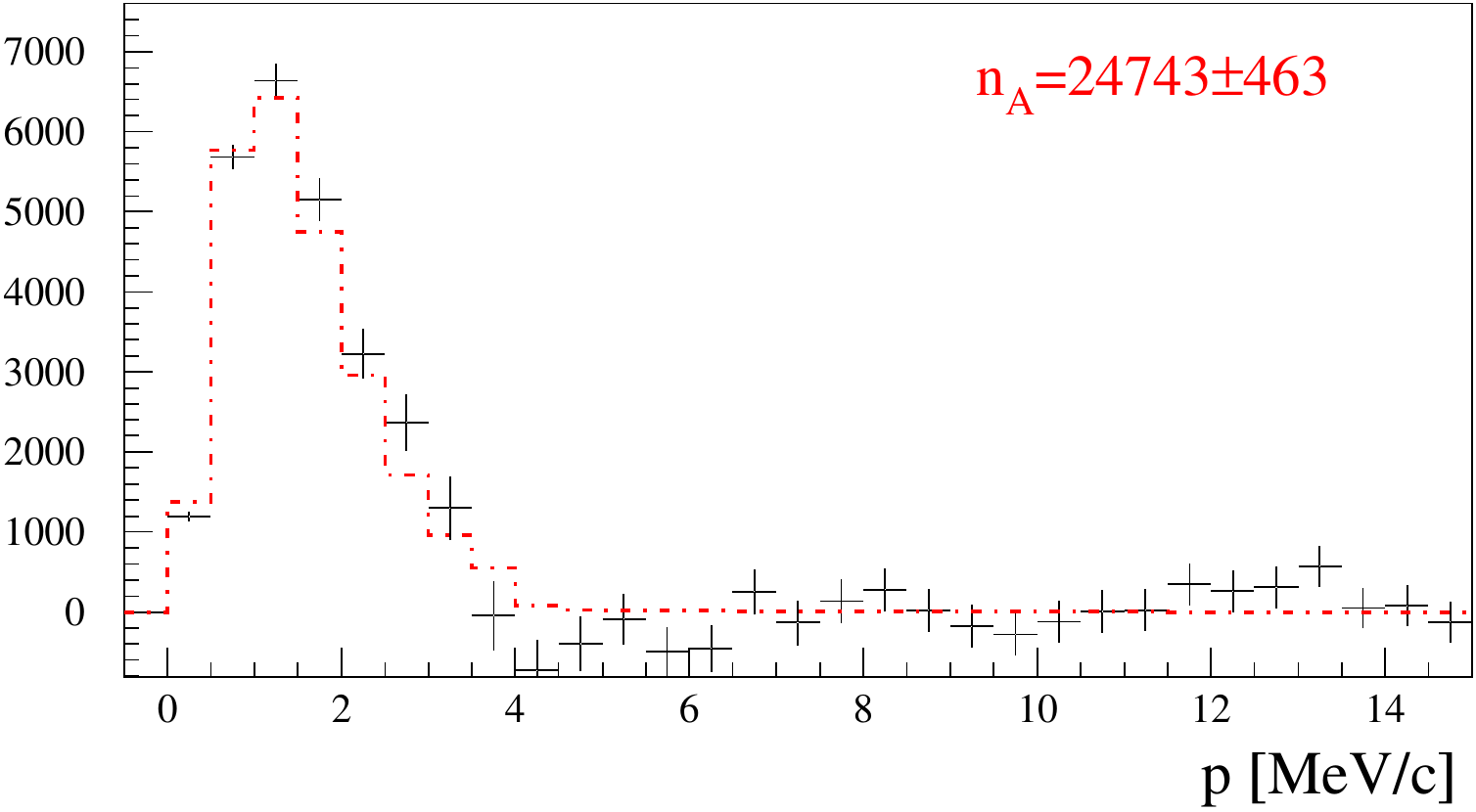}%
\caption{Typical relative momentum distributions of atomic pairs from $\pi^+\pi^-$ atom breakup detected in DIRAC experiment. Experimental data are shown as dots with errors bar (cross). The simulated data are show with red  dotted-dashed line. The number of detected atomic pairs in this distribution is $n_A=24743\pm463$.
\label{fig:qexp}}
\end{center}
\end{figure}

Significance of the difference in widths of relative momentum spectra calculated in different approaches can be evaluated at compression of these spectra with the experimental one.
A typical relative momentum spectrum of $\pi^+\pi^-$ atomic pairs detected in the DIRAC experiment is shown in Fig.~\ref{fig:qexp}, which is a mixture of pairs broken from different atomic states, dominantly with small $n$ and then detected. The simulated spectrum accounts numerically the population of atomic states, the relative momentum spectra of atomic pairs at breakup points, the particle multiple scattering of in the target and the resolution of detectors \cite{DIPGEN}. Considering that contributions of above-listed processes to the total spectrum width are summed quadratically we can made few conclusions. First, the relative momentum spectra at breakup points are significantly widen in the target and detectors. Nevertheless their contributions to the total width is not negligible, at least for few states with small $n$. Thus these spectra should be accurately accounted at calculation of the experimental number of atomic pairs. Second, the calculated spectra such as shown in Figs.~\ref{fig:1s-10s}--\ref{fig:q2s} can not be never observed experimentally and the detailed difference between the Born and Glauber spectra can not be checked. Third, for some simplified estimation even the accuracy of dipole approach is enough.

\medskip

\section{Summary}

The production of oppositely charged meson pairs in the breakup of relativistic EA interacting with matter has been considered. Making use the general expression for the creation amplitude of arbitrary continuous states, we obtain the analytical expressions for breakup spectra in the dipole and Born approximations. This allows to calculate the ionization spectra of EA scattered on screened Coulomb potential for arbitrary initial states of EA and thus to investigate dependences on all quantum numbers of initial states.

We considered the multiple photon exchanges for breakup of the ground state of EA at scattering on the unscreened Coulomb potential and obtained the closed analytical expression for relevant spectra. Exploiting the fact that multiple photon exchanges is important at small impact parameters (large transfer momenta) where one can safely neglects the screening of the Coulomb potential, whereas the effect of screening is crucial for the Born term determined by large impact parameters, we obtain the analytical expression for the ground state of EA breakup cross section accounting for all multiple (Glauber) photon exchanges  in the interaction with the target as well as all Coulomb interactions in initial bound state and final free meson pair. Using the obtained expressions we calculated the relevant spectra in different approaches.

\section*{Appendix A}\label{sec:app}

Let us derive the ionization form factor $S_{fi}(\vec q)$ defined by the equation
\begin{equation} S_{fi}(\vec q)=\int \psi_f^{\ast}
(\vec r)e^{i\vec q\vec r} \psi_i(\vec r)d^3r
\label{e1}
\end{equation}
for the transferred momentum $\vec q$, the EA wave function of the final (continuum) state,
which must be choose as
\begin{eqnarray}\label{e3}
\psi_f(\vec r)&=&\psi_{\vec p}^{(-)}=
c^{(-)}\exp(i\vec p\vec r)
\cdot\Phi\left[-i\xi,1,-i(pr+\vec p\vec r)\right]\,,\nonumber\\
c^{(-)}&=&(2\pi)^{-\frac{3}{2}}\exp\left(\frac{\pi\xi}{2}\right)
\Gamma(1+i\xi)\,, \quad \xi=\mu\alpha/p
\end{eqnarray}
\cite{land}, and the EA wave functions of the arbitrary initial bound state
\begin{equation}
\label{351}
\psi_i(\vec r)=\psi_{nlm}(\vec r)= Y_{lm}
\left(\frac{\vec r}{r}\right)R_{nl}(r)\,,
\end{equation}
with the radial part
\begin{eqnarray}
&&R_{nl}(r)=\frac{2\omega^{\frac{3}{2}}}{\Gamma(2l+2)}
\left[\frac{\Gamma(n+l+1)}{n\Gamma(n-l)}\right]^{\frac{1}{2}}(2\omega r)^l\times
\nonumber\\
&&\qquad\times\Phi(-n+l+1,2l+2;2\omega r)\cdot
\exp(-\omega r)\nonumber\\
&&\qquad=2\omega^{\frac{3}{2}}
\left[\frac{\Gamma(n-l)}{n\Gamma(n+l+1)}\right]^{\frac{1}{2}}(2\omega r)^l\times
\nonumber\\
&&\qquad\quad \times L^{2l+1}_{n-l-1}(2\omega r)\cdot
\exp(-\omega r),\quad \omega=\mu\alpha/n,
\end{eqnarray}
according to \cite{sommer,land}. Here, $\Phi$ is the confluent hypergeometric function and
$L^{2l+1}_{n-l-1}$ are the associated Laguerre polynomials.

\newpage

Making use of the recurrence relations \cite{ryzh,abram}
\begin{equation}
L_k^{\lambda+1}(x)=\frac{1}{x}\left[(k+\lambda+1)L_{k-1}^{\lambda}(x)
-(k+1)L_{k}^{\lambda}(x)\right]
\label{e4}
\end{equation}
and the representation of the Laguerre polynomials
in terms of the generating function
\begin{equation}
L_k^{\lambda}(x)=\Delta_z^{(k)}\left[(1-z)^{-(\lambda+1)}\exp
\left(\frac{xz}{z-1}\right)\right]
\label{e5}\, ,
\end{equation}
with the operator $\Delta_z^{(k)}$
\begin{equation}
\Delta_z^{(k)}\left[f(z)\right]=\frac{1}{k!}
\left.\left(\frac{d^k}{dz^k}f(z)\right)\right\vert_{z=0}
\label{e6}\, ,
\end{equation}
we rewrite the radial part of initial state wave function in the form
\begin{eqnarray}
&&R_{nl}=\frac{\omega^{\frac{1}{2}}}{r}
\left[\frac{\Gamma(n-l)}{n\Gamma(n+l+1)}\right]^{\frac{1}{2}}\cdot
(2\omega r)^l\times
\nonumber\\
&&\qquad\times\left[(n+l)\Delta_z^{(n-l-2)}-(n-l)\Delta_z^{(n-l-1)}\right]\times
\nonumber\\
&&\qquad\times\left[(1-z)^{-(l+1)}\exp\left(-\omega(z)r\right)\right];
\nonumber\\
&&\qquad\omega(z)=\omega\cdot(1+z)(1-z)^{-1}\, , \label{eq11}
\end{eqnarray}
which is more convenient for the further calculations.\\
The transition form factor (\ref{e1}) may be represent as a linear combination of the quantities
\begin{eqnarray}
\label{eq:352}\
&&I_{lm}^{j}=\Delta_z^{(j)}\left[(1-z)^{-(2l+1)} J_{lm} (\vec q,\vec p,z) \right]\,,
\nonumber\\
&&\quad j=n-l-2,\;n-l-1\, ,\nonumber\\
&&J_{lm}(\vec q,\vec p,z)=\int \frac{d^3r}{r}Y_{lm} \left(\frac{\vec
r}{r}\right) \Phi\left[i\xi,1;i(pr+\vec p\vec r)\right]\times
\nonumber\\
&&\quad\times\exp[i(\vec q-\vec p)\vec r-\omega(z)r](2\omega r)^l
\cdot\exp\left[-\omega(z)r\right]\,.
\label{eq:353}
\end{eqnarray}

In order to calculate (\ref{eq:353}), it is useful to represent the
hypergeometric function from (\ref{351}) in the form
\begin{eqnarray}
&&\Phi\left[i\xi,1;i(pr+\vec p\vec r)\right]=-\frac{1}{2\pi i}
\oint\limits_{C}^{} (-t)^{i\xi-1}(1-t)^{-i\xi}\times
\nonumber\\
&&\quad\times\exp[i\cdot t(pr+\vec p\vec r)]dt\,.
\label{eq:354}
\end{eqnarray}
Using the following relations
\begin{eqnarray}
\label{eq:355}
&&\exp(i\vec\tau\vec r)=4\pi\sum_{l=0}^{\infty}\sum_{m=-l}^{l}
i^{l}Y_{lm}\left(\frac{\vec\tau}{\tau}\right)
Y^{\ast}_{lm}\left(\frac{\vec r}{r}\right)j_{l}(\tau r)\, ,\nonumber\\
&& \quad j_{l}(x)=\sqrt{\frac{\pi}{2x}}J_{l+\frac{1}{2}}(x)\,,
\end{eqnarray}
\begin{equation}
\label{eq:357}
\int\limits_{0}^{\infty}r^{l+\frac{1}{2}}J_{l+\frac{1}{2}}(\tau r)
e^{-\bar\omega\cdot r}dr= \frac{(2\tau)^{l+\frac{1}{2}}\Gamma(l+1)}
{\sqrt{\pi}(\tau^2+\bar \omega^2)^{l+1}}\,,
\end{equation}
where
\begin{equation}
\label{eq:358}
\vec\tau=\vec q-\vec p(1-t),\quad
\bar\omega=\omega(z)-ip\cdot t\,,
\end{equation}
we find after simple calculations
\begin{eqnarray}
&&J_{lm}(\vec q,\vec p,z)=-\frac{\Gamma(l+1)}{2\pi i}
\oint\limits_{C}^{ }dt(-t)^{i\xi-1}(1-t)^{-i\xi}\times \nonumber\\
&&\quad\times{4\pi(4i\omega)^lY_{lm}\left(\vec \tau/\tau\right)\tau^l}
{(\tau^2+\bar \omega^2)^{l+1}}\,,
\label{eq:359}
\end{eqnarray}
\begin{eqnarray}
\label{eq:362}
&&\tau^2+\bar \omega^2=a(1-t)+c\cdot t\,, \quad a=\omega^2(z)+\bm\Delta^2\,,
\nonumber\\
&&\quad c=\left[\omega(z)-ip\right]^2+q^2\,.
\end{eqnarray}
Further, according to \cite{war}, we get
\begin{eqnarray}
\label{eq:360}
&&\!\!\!\!\! Y_{lm}\left(\frac{\vec\tau}{\tau}\right)\tau^l =
\sum_{l_1=0}^{l}q^{l_1}(-p)^{l-l_1}(1-t)^{l-l_1}\times\nonumber\\
&&\!\!\!\!\!\times\!\left[\frac{4\pi\Gamma(2l+2)} {\Gamma(2l_1+2)\Gamma(2l-2l_1+2)}\right]^{\!\!\frac{1}{2}} \!\!\!\!\cdot\!
\left[Y_{l_1}\left(\frac{\vec q}{q}\right)\otimes
Y_{l-l_1}\left(\frac{\vec p}{p}\right)\right]_{lm},\nonumber
\end{eqnarray}
\begin{eqnarray}
\label{eq:361}
&&\!\!\!\!\! \left[Y_{l_1}\left(\frac{\vec q}{q}\right)\otimes
Y_{l-l_1}\left(\frac{\vec p}{p}\right)\right]_{lm} = \nonumber\\
&&\sum_{m_1+m_2=m}\!\!\!\!\!\!C^{lm}_{l_1m_1(l-l_1)m_2}
Y_{l_1m_1}\left(\frac{\vec q}{q}\right)\cdot
Y_{(l-l_1)m_2}\left(\frac{\vec p}{p}\right).
\end{eqnarray}

Taking into account (\ref{eq:362}) and (\ref{eq:360}), it is easy to
see that (\ref{eq:359}) is the superposition of the quantities

\begin{eqnarray}
&&\!\!\!\!\! -\frac{1}{2\pi i}\oint\limits_{C}^{}
\frac{t^{i\xi-1}(1-t)^{-i\xi+l-l_1}} {[a(1-t)+ct]^{l+1}}=\nonumber\\
&&\!\!\!\!\! =a^{-(l+1)}\frac{\Gamma(1-i\xi+l-l_1)}{\Gamma(1-i\xi)}
F\left(i\xi,l+1;l-l_1+1;1-c/a\right)\nonumber\\
&&\!\!\!\!\! =a^{i\xi-l-1}c^{-i\xi}\frac{\Gamma(1-i\xi+l-l_1)} {\Gamma(1-i\xi)}
\times\nonumber\\
&&\quad\times F\left(i\xi,-l_1;l-l_1+1;1-a/c\right)\nonumber\\
&&\!\!\!\!\! =a^{i\xi-l-1}c^{-i\xi}\frac{\Gamma(l-l_1+1)\Gamma(l+1-i\xi)}
{\Gamma(l+1)\Gamma(1-i\xi)}
\times\nonumber\\
&&\quad\times F(i\xi,-l_1;i\xi-l;a/c)\nonumber\\
&&\!\!\!\!\! =\sum_{s=0}^{l_1}(-1)^{l-s} \frac{\Gamma(i\xi+s)\Gamma(l_1+1)}
{\Gamma(l_1-s+1)\Gamma(i\xi-l+s)\Gamma(s+1)\Gamma(l+1)}
\times\nonumber\\
&&\qquad\qquad\times a^{i\xi+s-l-1}c^{-s-i\xi}\nonumber\\
&&\!\!\!\!\! =(1-z)^{2l+2}\sum_{s=0}^{l_1}(-1)^{l-s} \times
\nonumber\\
&&\quad\times\frac{\Gamma(i\xi+s)\Gamma(l_1+1)}
{\Gamma(l_1-s+1)\Gamma(i\xi-l+s)\Gamma(s+1)\Gamma(l+1)}\times\nonumber\\
\label{eq:363}
&&\quad\times D_1^{i\xi+s-l-1}D_2^{-s-i\xi}\,,
\end{eqnarray}
where $D_{1,2}$ are given by
\begin{eqnarray}
D_1&=&(1+z^2)(\omega^2+\Delta^2)-2z(\Delta^2-\omega^2)\,, \nonumber\\
D_2&=&(\omega-ip)^2+q^2-2z(q^2-p^2-\omega^2)+ \nonumber\\
&&+z^2[(\omega+ip)^2+q^2]\,.\label{eq3}
\end{eqnarray}

The further calculations are the same as in \cite{bv}.
Omitting the simple but cumbersome algebra, we finally obtain
\begin{eqnarray}
&&\!\!\!\!\! S_{\vec p,nlm}(\vec q)=4\pi\cdot 2^{2l}i^l\omega^{l+\frac{1}{2}}
\left[\frac{\Gamma(n-l)}{n\Gamma(n+l+1)}\right]^{\frac{1}{2}}\times\nonumber\\
&&\times\sum_{s=0}^{l}G_{lms}(\vec p,\vec q)H_{nls}(\vec p,\vec q)
(\omega^2+\Delta ^2)^{i\xi+s-l-1} \times\nonumber\\
&&\times[(\omega-ip)^2+q^2]^{-s-i\xi}\,,\nonumber\\
\label{eq:365}
&&\!\!\!\!\! G_{lms}(\vec p,\vec q)=(-1)^{l-s}
\frac{\Gamma(i\xi+s)}{\Gamma(i\xi-l+s)\Gamma(s+1)}\nonumber\\
&&\times\sum_{l_1=s}^{l}\left[\frac{4\pi\Gamma(2l+2)}{\Gamma(2l_1+2)\Gamma(2l-2l_1+2)}\right]^{\frac{1}{2}}\times \nonumber\\
&&\times\frac{\Gamma(l_1+1)}{\Gamma(l_1-s+1)}q^{l_1}(-p)^{l-l_1}
\times\left[Y_{l_1}\left(\frac{\vec q}{q}\right)\otimes
Y_{l-l_1}\left(\frac{\vec p}{p}\right)\right]_{lm}\, ,\nonumber\\
\label{eq:366}
&&\!\!\!\!\! H_{nls}(\vec p,\vec q)=(n+l)F_{n_1ls}(\vec p,\vec
q) - (n-l)F_{n_2ls}(\vec p,\vec q)\, ,\nonumber\\
&&n_1=n-l-1\, ,\quad n_2=n-l-2\, ,\nonumber\\
\label{eq:367}
&&\!\!\!\!\! F_{n_{1(2)}ls}(\vec p,\vec q)=
\frac{\Gamma(l-s+\frac{1}{2}-i\xi)}{\Gamma(2l-2s+1-2i\xi)}\times
\sum_{k=0}^{n_{1(2)}}w^kC_k^{(i\xi+s)}(v)\times\nonumber\\
&&\times\frac{\Gamma(n_{1(2)}-k+2l-2s+1-2i\xi)}{\Gamma(n_{1(2)}-k+l-s+\frac{1}{2}-i\xi)}\times \nonumber\\
&&\times P_{n_{1(2)}-k}^{(l-s-\frac{1}{2}-i\xi,l-s+\frac{1}{2}-i\xi)}(u)\, .
\end{eqnarray}

\end{document}